\documentclass[aip,reprint]{revtex4-1}

\usepackage{graphicx}
\usepackage{dcolumn}
\usepackage{bm}
\usepackage{amsmath,amsfonts,amsthm}
\usepackage[table,xcdraw]{xcolor}
\usepackage{multirow}

\draft 

\begin{document}


\title{Stabilization of coupled Dzyaloshinskii domain walls in fully compensated synthetic anti-ferromagnets} 



\author{Nisrit Pandey}%
 \email{npandey@andrew.cmu.edu}
 \affiliation{Department of Materials Science \& Engineering, Carnegie Mellon University,\\ Pittsburgh, PA 15213 USA.}

\author{Maxwell Li}%
 \affiliation{Department of Materials Science \& Engineering, Carnegie Mellon University,\\ Pittsburgh, PA 15213 USA.}

\author{Marc De Graef}%
 \affiliation{Department of Materials Science \& Engineering, Carnegie Mellon University,\\ Pittsburgh, PA 15213 USA.}

\author{Vincent Sokalski}%
 \affiliation{Department of Materials Science \& Engineering, Carnegie Mellon University,\\ Pittsburgh, PA 15213 USA.}


\date{\today}

\begin{abstract}

We examine the combined effects of interlayer exchange coupling (IEC) and the interfacial Dzyaloshinskii-Moriya Interaction (DMI) on the structure of magnetic domain walls in fully compensated synthetic anti-ferromagnets (SAFs).  Ir-based SAFs with ferromagnetic (FM) layers based on [Pt/(Co/Ni)$_M$]$_N$ were characterized by Lorentz transmission electron microscopy (LTEM).  The multi-layer design of the individual ferromagnetic layers enables control of the interfacial Dzyaloshinskii-Moriya interaction (via ‘M’) and, in turn, the structure and chirality of domain walls (DWs). We compare the Fresnel-mode LTEM images in SAF designs with only a change in the purported strength of the DMI.  The existence of anti-ferromagnetically coupled Dzyaloshinskii domain walls (DWs) in a high DMI SAF is confirmed through application of in-situ perpendicular magnetic field and sample tilt.  This conclusion is based on a unique set of conditions required to observe contrast in Fresnel-mode LTEM, which we outline in this document.  

\end{abstract}


\maketitle 

\section{\label{sec:level1}Introduction}

Synthetic anti-ferromagnets (SAFs) have had a profound impact on magnetic storage technology over the course of the past several decades, particularly, for their role in magnetoresistive read heads.\cite{Edwards1991,vandenBerg1996,Inomata2002}  There has been a striking re-emergence of interest in SAFs, recently, for their potential use in chiral spintronic applications as well.\cite{Yang2015B,Zhang2016,Tomasello2017,Prudnikov2018,Legrand2019} This is due to the now well-established interfacial Dzyaloshinskii-Moriya interaction\cite{Dzyaloshinsky1958,Moriya1960,Thiaville2012} (DMI), which leads to chiral N\'eel domain walls (DWs), called Dzyaloshinskii DWs, or Skyrmions in thin films with broken mirror symmetry.\cite{Chen2013,Franken2014,Lau2016,MoreauLuchaire2016,McVitie2018,Li2019}  Significant improvement in the current-induced velocity of these features when incorporated into SAFs has been identified when driven by spin-orbit torques (e.g. the spin Hall effect).\cite{Liu2012,Yang2015B}  However, magnetically imaging the detailed structure of these features in SAFs is notoriously difficult especially when the ferromagnetic layers are fully compensated, which is incidentally when peak performance is expected.\cite{Yang2015B}  Moreover, new materials challenges emerge when trying to combine multiple interface phenomena into a single heterostructure.  

\begin{figure}[b]
\includegraphics[width = \columnwidth]{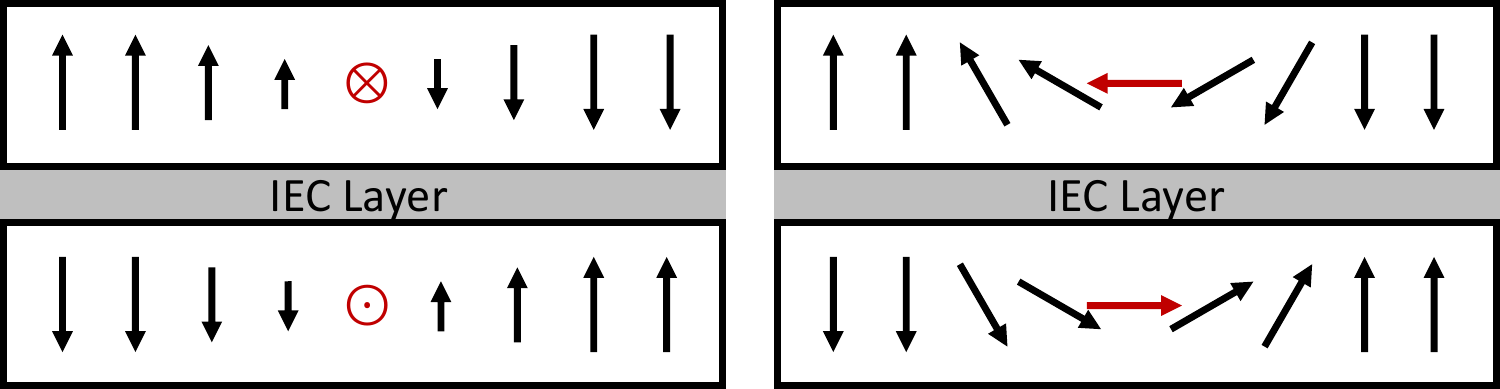}%
\caption{\label{SAF_Types} Expected structure of synthetic antiferromagnetic domain walls for the case of weak (left) and strong (right) interfacial DMI. The weak case should result in coupled, achiral Bloch DWs.  The strong case (where the sign of DMI is the same in both layers) should result in coupled Dzyaloshinskii DWs with antiparallel internal magnetization.}%
\end{figure}

Here, we present a tunable SAF design that leverages the IEC of Co/Ir/Co, the interfacial DMI of Pt/Co, and the perpendicular magnetic anisotropy (PMA) of Co/Ni.  Our film design (as described in a subsequent section) allows us to independently tune the former two without compromising the latter.  The structure of coupled DWs in our fully compensated SAF (see Figure \ref{SAF_Types}) is determined from in-situ Fresnel-mode Lorentz TEM based on a combination of field and tilt conditions under which DW contrast is observed. 
 
\begin{table}[h]
\centering
\begin{tabular}{|c|c||c|c|c|c|}
\hline
\multicolumn{2}{|c||}{Imaging Conditions} & \multicolumn{4}{c|}{Contrast Expected} \\ \hline
 &  & \multicolumn{2}{c|}{Single Layer} & \multicolumn{2}{c|}{SAF} \\ \cline{3-6} 
\multirow{-2}{*}{\begin{tabular}[c]{@{}c@{}}Specimen \\ Tilted\end{tabular}} & \multirow{-2}{*}{\begin{tabular}[c]{@{}c@{}}Perp. Field \\ Applied\end{tabular}} & Bloch & N\'eel & Bloch & N\'eel \\ \hline
$\times$ & $\times$ & Yes & No & No & No \\
\checkmark & $\times$ & Yes & Yes & No & No \\
$\times$ & \checkmark & Yes & No & Yes & No \\
\checkmark &\checkmark & Yes & Yes & Yes & Yes \\ \hline
\end{tabular}%
\caption{Conditions under which Fresnel-mode contrast reveals domain walls in a single perpendicular ferromagnetic layer and a fully compensated synthetic antiferromagnet.}
\label{Conditions}
\end{table}

\section{Experimental Techniques}

The multi-layers examined in this work were deposited onto 10 nm thick amorphous Si$_3$N$_4$ TEM membranes via dc magnetron sputtering in an Ar environment with working pressure fixed at 2.5 mTorr and base pressure of $< 2.5 \times 10^{-7}$ Torr. The adhesion/seedlayer was always Ta(30)/Pt(30) to promote FCC(111) texture through the sample while a capping layer of Pt(30) was deposited for protection.  All units in parentheses are \AA.  Multi-layers of [Co(2)/Ni(6)]$_2$/Co(2)/Ir(0-25)/[Co(2)/Ni(6)]$_2$/Co(2) were grown with an aperture-based wedge growth technique that varies the thickness of Ir to measure its impact on the interlayer exchange coupling strength, $\sigma_\text{exc}$, as determined from $\sigma_\text{exc}=\mu_0H_\text{exc}M_\text{s}t_\text{f}$, where $H_\text{exc}$ is the exchange coupling field measured from an MH loop, $t_\text{f}$ is FM thickness and $M_\text{s}$ ($=600$ kA/m) is the saturation magnetization.  The SAF designs for examination by LTEM were based on the following structure: \{Pt(5)/[Co(2)/Ni(6)]$_M$\}$_N$/Co(2)/Ir(3) /\{[Co(2)/(Ni(6)]$_M$/Pt(5)\}$_N$.\footnote{To balance the magnetization of the two layers an additional Co(2) layer was included in the top ferromagnet as indicated in the insets of the respective sketches of figure \ref{SAF}}  The stacking sequence is such that the same sign of interfacial DMI will result for both the top and bottom ferromagnets. It is expected that a reduction in M will lead to an increase in the DMI strength while an increase in N should only affect the strength of IEC.  Moreover, higher N will increase the total magnetic induction of the specimen leading to improved resolution.  In this paper we consider two cases: M=6, N=1 and M=1, N=6, which we refer to, henceforth, as the ``weak" and ``strong" DMI SAFs, respectively.  An additional sample of the form \{[Co(2)/Ni(6)]$_1$/Pt(5)\}$_{50}$ was deposited to independently characterize the DW structure of what is expected to be a high DMI sample. 

Fresnel-mode LTEM imaging was performed at room temperature on an aberration-corrected FEI Titan G2 80-300 with an accelerating voltage of $300$ kV. A perpendicular magnetic field was applied \textit{in situ} by weakly exciting the objective lens of the microscope. This method of magnetic imaging forms contrast out of the focus plane through the deflection of electrons via the Lorentz force.\cite{degraef2000d} 
\begin{figure}[h]
\includegraphics{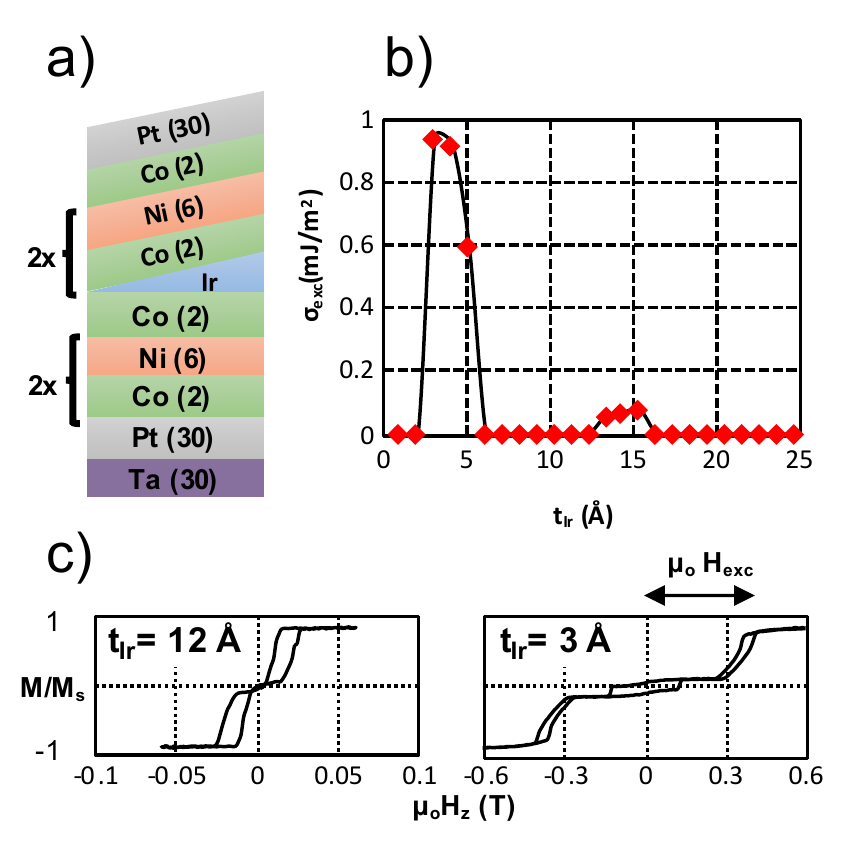}%
\caption{\label{MH} a) schematic of the wedge film stack used to evaluate interlayer exchange coupling through Co/Ir/Co. b) $\sigma_\text{exc}$ vs. t$_{Ir}$ determined from the film stack in (a). c) representative perpendicular M-H loops for $t_\text{Ir} = 3$ \AA\  and $12$ \AA.}%
\end{figure}

\section{Results and Discussion}

The results of interlayer exchange coupling as a function of the Ir layer thickness in [Pt/(Co/Ni)$_6$]/Co/Ir(t$_{Ir}$)/Co/[(Ni/Co)$_2$/Pt] multi-layers are shown in Figure~\ref{MH}.  Peaks in the anti-ferromagnetic coupling (defined herein as positive) were observed between 2-6 \AA\ (strong) and 12-16 \AA\ (weak). The strongest IEC was detected at $t_\text{Ir}=3$ \AA, which was used as the spacer layer thickness for subsequent SAF stacks. 

\begin{figure}[t]
\includegraphics[width = 2in]{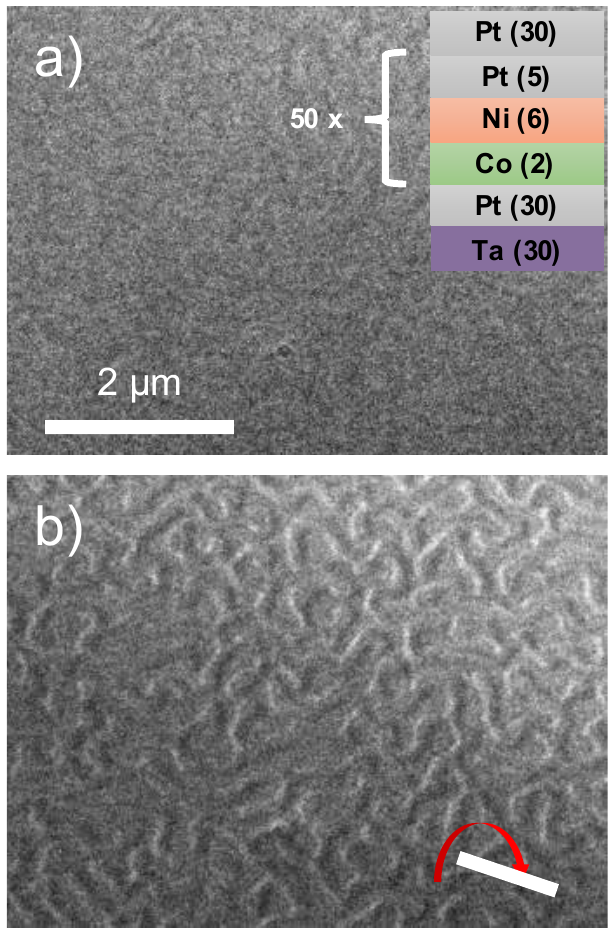}%
\caption{\label{PtCoNi}Fresnel-mode Lorentz TEM images of [Pt/Co/Ni]$_{50}$ in a) the absence of and b) in the presence of a 20$^\circ$ sample tilt. Inset depicts schematic of multi-layer film stack.}%
\end{figure}

\begin{figure*}[t]
\includegraphics{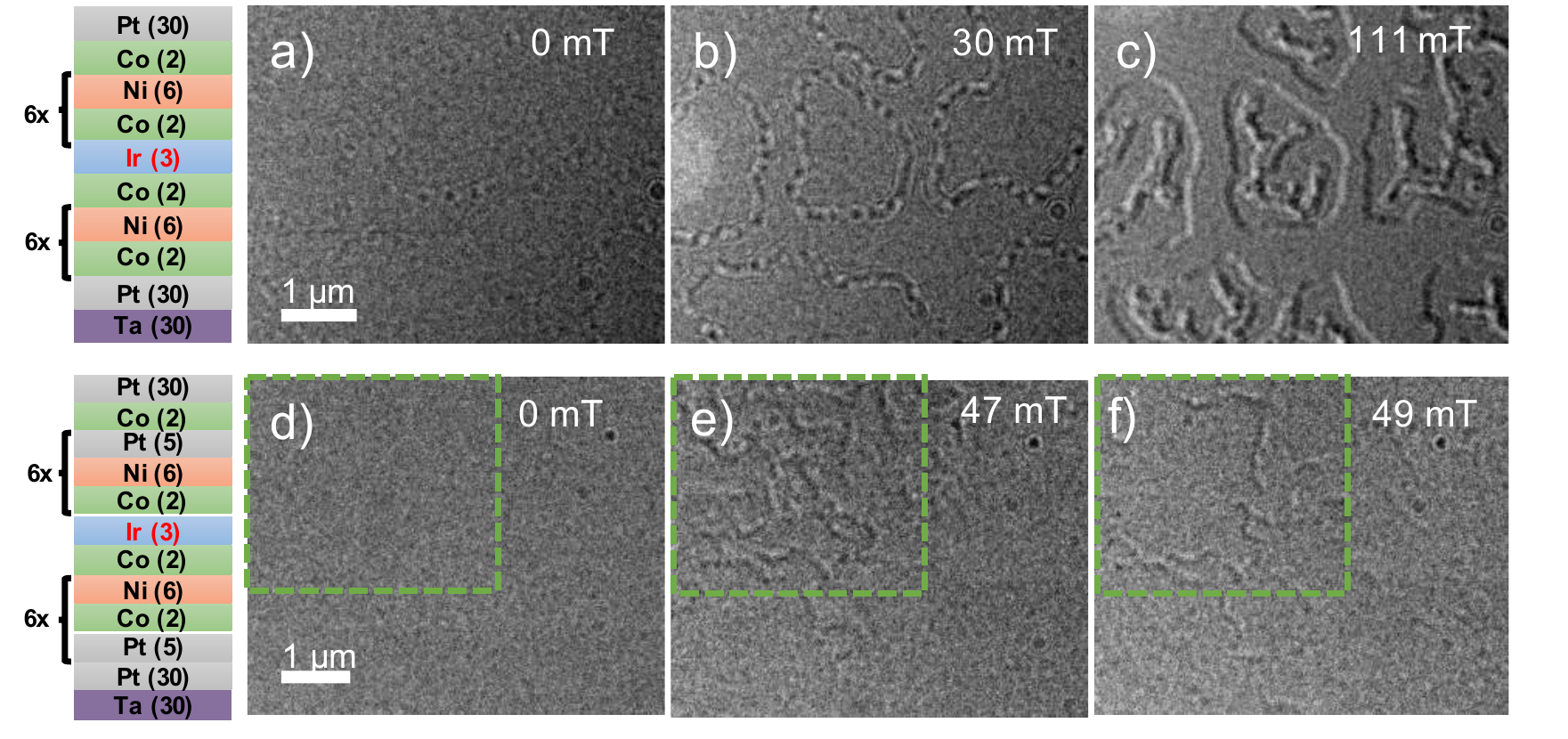}%
\caption{\label{SAF}Fresnel-mode Lorentz TEM images of the SAF with a-c) ``weak" and d-f) ``strong" DMI in the presence of increasing perpendicular magnetic field applied \textit{in situ}. We note that a sample tilt was not applied in these images; magnetic contrast in the ``strong" DMI sample coincides with regions where bends in the TEM membrane are observed and are highlighted by the dotted lines. Schematics of the multi-layer schemes are also depicted.}%
\end{figure*}

Before proceeding, we note some important features of Fresnel-mode LTEM imaging as it pertains to SAFs with PMA.  Perpendicular magnetic domains only contribute to the image contrast if the specimen is tilted as governed by the Lorentz force.  For domain walls, only the Bloch component produces contrast.  Consequently, the existence of N\'eel DWs is confirmed by the absence (presence) of contrast in the untilted (tilted) specimen.  The situation becomes more complicated for the case of fully compensated SAFs as outlined in Table \ref{Conditions}.  Notably, contrast should be absent for the case of the top and bottom domain walls in perfect alignment even if the specimen is tilted regardless of the DW structure.  For the case of Bloch DWs, the application of a perpendicular field should drive the DWs apart and their contrast would be revealed.  For the case of N\'eel DWs, even though the field drives the walls apart, they have no contrast signature without specimen tilt.  Therefore, confirmation of N\'eel DWs in a fully compensated SAF requires the absence of contrast in all cases except for the combination of field and tilt as outlined in Table \ref{Conditions}.

Figure~\ref{PtCoNi} shows Fresnel-mode images of the sample with a single perpendicular magnetic layer (\{[Co(2)/Ni(6)]$_1$/Pt(5)\}$_{50}$) with and without sample tilt.  The observation of Fresnel-mode contrast only in the tilted orientation confirms the N\'eel structure of the DWs. Some comments on the origin of DMI in this ternary superlattice are in order.  It is widely noted that the DMI originates from the spin-orbit interaction, in this case associated with heavy transition metals.\cite{Yang2015}  However, it is evident that the interfacial DMI is, naturally, a property of the interface and should not be expected to be the same in comparing Pt/Co vs Ni/Pt.  The results shown here and previous measurements of the DMI in a Pt/(Co/Ni)$_2$/Pt heterostructure by Kerr microscopy confirm this.\cite{LauThesis2018}  Therefore, the requirement of sandwiching a ferromagnetic sample by two different heavy metals can be avoided if the magnetic material itself at the two interfaces is different.  A similar observation has been made previously for the case of Ir/Co/Ni based ternary superlattices imaged by spin-polarized low energy electron microscopy (SPLEEM).\cite{chen2015ternary}


We now move on to LTEM of the two SAFs under consideration. In the zero-field condition (with or without tilt) for the ``weak" DMI SAF, a small amount of contrast is observed at regions later confirmed to be DWs. This alternating contrast, which varies along the length of the DW, is referred to as a ``tiger-tail" and corresponds to a slight relative displacement between the top and bottom DWs associated with dipolar interactions.\cite{Kiseleva2007,Kiseleva2010}  Because this contrast did not change significantly with sample tilt, we confirm that the effective magnetization of the top and bottom layers are roughly compensating.  When a perpendicular magnetic field is applied, the ``tiger-tail" contrast becomes more pronounced as some of the partially coupled DWs are driven apart.  Additional increases in field result in wider separation of DWs in top and bottom layer (Figure \ref{SAF}c)), which eventually annihilate before saturation.  

To examine the strong DMI SAF, we image the specimen in close proximity to a bend in the Si$_3$N$_4$ membrane.  Such bend contours (up to 15$^\circ$) are common in these Si$_3$N$_4$ TEM membranes, especially along the edges of the membrane window.\cite{Fallon2019}  This allows us to simultaneously view tilted and untilted areas of the specimen.  In Figure \ref{SAF}d-f), the upper half of the field of view is tilted with respect to the electron beam.  We note that no contrast is observed anywhere in the field of view without the applied field.  This observation indicates that the top and bottom DWs are directly aligned unlike the weak DMI sample.  Such a result is likely due to the dipolar interactions between the DWs themselves where the anti-parallel alignment of the internal magnetization results in stray field closure.  The only condition where the strong DMI SAF reveals DW contrast is in the tilted region (Figure \ref{SAF}e-f)) when a perpendicular magnetic field is applied.  In comparison with Table \ref{Conditions}, this corresponds to the case of pure N\'eel DWs in a fully compensated SAF. As with the weak DMI SAF, the DWs are driven apart with application of subsequently larger fields prior to saturation. 

\section{Summary}

We have examined the structure of DWs in fully compensated Ir-based SAFs with weak and strong DMI as controlled by the multi-layer design of the Pt/Co/Ni based ferromagnetic layers.  We first demonstrate that Dzyaloshinskii DWs are stabilized in the single ferromagnetic layer, which confirms that a significant DMI results from the combination of Pt/Co and Ni/Pt interfaces in the superlattice.  We then conclude that the strong DMI SAF adopts coupled Dzyaloshinskii DWs based on Fresnel-mode LTEM.  This is based on the unique requirement that the specimen is both subject to a perpendicular field to drive the DWs apart and a specimen tilt to reveal contrast associated with the N\'eel configuration.  With interest growing in chiral SAFs where unwanted stray fields and/or the Skyrmion Hall effect cancel, we are hopeful that the tunable material combinations and approach to LTEM imaging presented in this work will be adopted for future investigations.


%
%

\begin{acknowledgments}
This work is financially supported by the Defense Advanced Research Project Agency (DARPA) program on Topological Excitations in Electronics (TEE) under grant number D18AP00011. The authors also acknowledge use of the Materials Characterization Facility at Carnegie Mellon University supported by grant MCF-677785.
\end{acknowledgments}

\bibliography{aipsamp}

\end{document}